\documentclass[aps,twocolumn,
showpacs,tightenlines]{revtex4}
\usepackage{amsfonts,amssymb,amsmath}

\def\({\left(}
\def\){\right)}

\newcommand{\beq}{\begin{eqnarray}}
\newcommand{\eeq}{\end{eqnarray}}

\def\bce{\begin{center}}
\def\ece{\end{center}}

\newcommand{\be}{\begin{equation}}
\newcommand{\ee}{\end{equation}}
\newcommand{\bea}{\begin{eqnarray}}
\newcommand{\eea}{\end{eqnarray}}
\newcommand{\beaa}{\begin{eqnarray*}}
\newcommand{\eeaa}{\end{eqnarray*}}

\def\treug{\langle\bar{\psi}\psi\rangle}
\newcommand{\bb}{\begin{equation}}

\begin{document}

\title{Accelerating and decelerating cosmology from spinor and scalar 
fields non-minimally coupled with f(R) gravity}


\author{Yu.A. Rybalov, A. N. Makarenko, K.E. Osetrin} 
\affiliation{Department of Theoretical Physics, Tomsk State Pedagogical 
University, Tomsk, 634041, Russia.}
\begin{abstract}
In this paper we investigate the accelerating and decelerating 
cosmological models with 
non-linear spinor fields and non-minimal interaction of $f(R)$ gravity with a 
scalar field. We combine
  two different approaches to the description of dark energy: modified gravity 
theory and introduction of the additional fields.
Solutions for the  FRW universe with power-law
scale factor  are reconstructed for the model under consideration with specific 
choice for scalar and spinor potentials.
It is explained the role of scalar and spinor potentials as well as f(R) 
function for emergence of accelerating or decelerating cosmology.
  \end{abstract}
\keywords{Dark energy, spinor fields, scalar fields, non-minimal interraction}

\maketitle

\section{Introduction}

The problem of the dark
energy and dark matter
is one of the main
challenges of modern cosmology.
Astrophysical data indicate that the observed universe is  in an
accelerated phase  \cite{dat0}.
This acceleration could be  induced
by  the so-called dark energy
(see Ref. \cite{review} for a recent review).
On the other hand, astrophysical observations provide evidence \cite{dm1} for 
the existence
of a non-baryonic, non-interacting and pressure-less component of the Universe, 
dubbed dark
matter.  This leads us to the need to revise the standard cosmology.

The  cosmological constant models are the simplest candidates for the solution 
of the problem of the universe acceleration. However, these models have still 
problems with the consistent description of the different evolution stages of 
the Universe. Scalar theory is most popular to describe the current 
accelerating expansion and early-time inflation. However, to describe the dark 
matter we have to introduce additional fields.  One can consider a model with 
two scalar fields \cite{LM1} (or scalar field and lagrange multiplier(s) 
\cite{LM11}), or, for example, models with additional spinor field to describe 
dark energy and dark matter.

The study of spinor fields in curved spacetime has a long history.  The Dirac 
equation was investigated for massless spinor fields in
curved space-time more than 50 years ago \cite{Brill:1957fx}.  Spinor fields 
can be used to describe the  primordial  inflation \cite{Armendariz-Picon:2003} 
and current expansion  \cite{Ribas:2005}. However, the exact solutions in the presence of the spinor field is difficult to build (for example, see  \cite{Mak_Os}).

A significant number of attempts have been made to construct the cosmological 
models with a spinor field for description of dark energy, where a 
non-canonical kinetic term was considered, such as k-inflation and k-essence 
models \cite{lit17}. In \cite{lit18}, the properties of one of the foregoing 
models with self-interacting spinor  with the noncanonical kinetic term were 
studied. 

It should be noted that the models involving the squared classical Dirac 
Lagrangian can be considered as a special case of the k-essence model 
\cite{lit19}. The scalar invariant constructed from two spinor fields 
dynamically develops  a nonvanishing value in Quantum Chromodynamics (QCD) 
theory \cite{lit20}. In this case, the chiral symmetry for the spinor field is 
broken. This is of significance for the evolution of the Universe. Only in a 
few papers the dynamic symmetry was assessed with the aim of interrupting 
non-static behavior of spacetime.

There is another way to solve the problem of dark energy that does not require 
the introduction of the dark component.
The modified theory of gravity may be quite realistic to describe the different 
phase of evolution of the Universe (see recent review \cite{lit7}).
A simple model describing the unified description for primordial inflation and 
current accelerating expansion was presented by Nojiri-Odintsov in \cite{lit8}. 
It also shows the viability of the modified gravity \cite{lit9, lit10, lit11, 
lit12}, which describes the $\Lambda CDM$ epoch, like the standard theory with 
cosmological constant. In addition,  such models satisfy the tests of solar 
system (see \cite{lit13}), as well as to adequately describe all the stages of 
development of the universe, starting with the early inflation until late 
accelerated expansion, with a correct description of the intermediate stage 
\cite{lit10, lit11}.

In order to merge two approaches, we could use non-local gravity  
and present its
local (scalar-tensor) formulation.
Such theories also  naturally lead to the unification of inflation with 
late-time cosmic acceleration.

In the recent paper we considered a cosmological model with a spinor field and 
a scalar field couples with an arbitrary function of  the curvature.
  Of course, such models are not standard ones, in the sense that they are not 
multiplicatively
renormalizable in curved spacetime \cite{Buch}. Hence, they should be 
considered as kind of
effective theories (without clear understanding of their origin and their 
relation with
more fundamental string/M-theory).
The present paper is devoted to study of non-minimally coupled scalar theory 
introduced in refs.\cite{lit2} with self-interacting spinor field.
We study the FRW equations of motion for such non-linear and non-minimal system 
with scalar and spinor fields. Specific choice of scalar and spinor potentials 
is made in the process of the search of explicit accelerating/decelerating 
cosmological 
solutions. Several power-law solutions for current dark energy epoch are 
constructed. It is known that these cosmologies are quite realistic and pass 
the observational bounds.

\section{The field equations}

Let us consider a model with the action in the form:
\bb S=\int{d^4\sqrt{-g}\left(\frac{R}{2}+f(R){\cal L}_{\phi}+{\cal L}_D\right) } , 
\label{act}\ee
where $R$ is scalar curvature.

The Lagrangian of a scalar field of mass $m$ is given by: \bb
{\cal
L}_{\phi}=\frac{1}{2}g^{\mu\nu}\phi_{;\;\mu}\phi_{;\;\nu}-V(\phi),\label{lagphi}
\ee where $V(\phi)$ is the potential of the scalar field. The Dirac Lagrangian
${\cal L}_D$ of fermion field of mass $m_f$ has the  form: \bb {\cal
L}_D=\frac{i}{2}\{\bar{\psi}\Gamma^{\mu}D_{\mu}\psi-D_{\mu}\bar{\psi}\Gamma^{\mu}\psi\}-m_f\treug-V(\psi\bar\psi)
. \label{dir}\ee
In the expression (\ref{dir}), $V(\psi\bar\psi)$
describes the potential of fermion field and
$\bar{\psi}=\psi^\dag\gamma^0$ denotes the conjugate spinor.
$\Gamma^{\mu}=e^{\mu}_{a}\gamma^a$ are generalized  the Dirac matrices  in a 
curved spacetim ($e^{\mu}_{a}$ is tetrad). The covariant derivative in the 
equation (\ref{dir}) is defined by the rule:
\begin{eqnarray}
D_{\mu}\psi&=&\partial_{\mu}\psi-\Omega_{\mu}\psi,\nonumber\\
D_{\mu}\bar{\psi}&=&\partial_{\mu}\bar{\psi}-\bar\psi\Omega_{\mu},
\end{eqnarray}
where
\bb
\Omega_\mu=-\frac{1}{4}g_{\rho\sigma}\left[\Gamma_{\mu\delta}^{\rho}
-e^{\rho}_{b}\partial_{\mu}e^{b}_{\delta}\right]\Gamma^{\sigma}\Gamma^{\delta},
\ee
Here $\Gamma^\rho_{\mu\sigma}$ are Christoffel symbols.

Let us now consider  a Friedmann-Robetson-Walker (FRW)  universe with the
flat spatial metric
\bb ds^2=dt^2-a(t)^2(dx^2+dy^2+dz^2). \label{met}\ee
From Eq. (\ref{act}) and (\ref{dir}) one can obtain  equation for the spinor 
field
\bb \frac{dL_D}{d\bar\psi}=\frac{dL_D}{d\psi}=0 \label{modir}\ee
or
\begin{eqnarray}
i(D_{\mu}\bar\psi)\Gamma^\mu+m_f\bar\psi+\frac{dV}{d\psi}&=&0,\\
i\Gamma^\mu D_{\mu}\psi-m_f\psi-\frac{dV}{d\bar\psi}&=&0.\label{Dirac1}
\end{eqnarray}

Einstein's equations can be written as
\bb R_{\mu\sigma}-\frac12g_{\mu\nu}R=-T_{\mu\nu},\label{en}\ee
where
$T_{\mu\nu}=(T_f)_{\mu\nu}+(T_\phi)_{\mu\nu},$
   ($T_f)_{\mu\nu}$ is the energy-momentum tensor of the fermion fields and
$(T_\phi)_{\mu\nu}$  is the contribution of the variation of the scalar field 
which non-minimally interacts with $F(R)$ . A symmetric form of the 
energy-momentum of the fermion field is as follows
  \bea &&(T_f)^{\mu\nu}=-g^{\mu\nu}L_D+\\
&&+\frac{i}{4}\{\bar\psi\Gamma^\mu
D^\nu\psi+\bar\psi\Gamma^\nu D^\mu\psi
-D^\nu\bar\psi\Gamma^\mu\psi-D^\mu\bar\psi\Gamma^\nu\psi\},\nonumber
\eea where $\Gamma^0=\gamma^0$,
$\Gamma^i=\frac{1}{a(t)}\gamma^i$, $\Gamma^5=\gamma^5$ and \bb
\Omega_0=0,\; \Omega_i=\frac12\dot
a(t)\gamma^i\gamma^0\label{omeg}.\ee From the equations (\ref{omeg}),
  (\ref{act}), (\ref{dir}) and (\ref{met}) we get the non-zero components of the 
energy-momentum temsor of the fermion field \bb
(T_f)^0_0=m_f\treug+V,\label{too1}\ee \bb
(T_f)^i_i=V-\frac{\bar\psi}2\frac{dV}{d\bar\psi}-\frac{dV}{d\psi}\frac{\psi}2.\label{tii1}\ee

The interaction between the fermionic components is modeled by a 
non-equilibrium pressure ($\varpi$) in the energy-momentum tensor source. 


A symmetric form the 
energy-momentum tensor of the scalar field can be obtained from (\ref{act}) in 
the  form
\begin{eqnarray}
&&(T_\phi)_{\mu\nu}=-({\cal 
L}_{\phi}f'R_{\mu\nu}+f\phi_{;\;\mu}\phi_{;\;\nu}-\\
&&-2\left[\Box{\cal L}_{\phi}f'+2{\cal L}_{\phi\; 
;\;\sigma}f'_{;\;\sigma}+{\cal L}_{\phi}\Box f'\right]g_{\mu\nu}+\nonumber\\
&&+2\left[{\cal L}_{\phi\; ;\; \mu\nu}f'+{\cal L}_{\phi\;
;\;\mu}f'_{;\;\nu}+{\cal L}_{\phi\; ;\;\nu}f'_{;\;\mu}+{\cal
L}_{\phi}f'_{;\;\mu\nu}\right]).\nonumber
\end{eqnarray}
\begin{eqnarray}
&&(T_\phi)^0_0=p_{\phi}=-({\cal L}_{\phi}f'R^0_0+f\dot{\phi}^2-\\
&&-2\left[\Box{\cal L}_{\phi}f'+2{\cal L}_{\phi\; 
;\;\sigma}f'_{;\;\sigma}+{\cal L}_{\phi}\Box f'\right]+\nonumber\\
&&+2\left[{\cal L}_{\phi\; ,\; 00}f'+{\cal L}_{\phi\;
,\;0}f'_{,\;0}+{\cal L}_{\phi\; ,\;0}f'_{,\;0}+{\cal
L}_{\phi}f'_{,\;00}\right]).\nonumber
\end{eqnarray}
\begin{eqnarray}
&&(T_\phi)^i_i=-\rho_{\phi}=\\
&&=-({\cal
L}_{\phi}f'R^i_i-2\left[\Box{\cal L}_{\phi}f'+2{\cal L}_{\phi\;
;\;\sigma}f'_{;\;\sigma}+{\cal L}_{\phi}\Box f'\right]).\nonumber
\end{eqnarray}


We now write the equation of motion of the scalar field as \bea \label{scal}
&&f(R)\Box\phi+g^{\mu\nu} f_{;\; \mu}\phi_{;\;\nu}+V'(\phi)f(R)=0,\nonumber\\
&&V'(\phi)=\frac{d V(\phi)}{d \phi}. \eea


The consequence of the equations of motion of the spinor fields is given by:
\bb \frac{d}{dt}\bar{\psi}\psi+3H\bar{\psi}\psi=0, \ee

or

\bb \frac{\dot{\bar{\psi}\psi}}{\bar{\psi}\psi}=-3H,
\label{eqpsipsib} \ee \bb \label{eqpsipsib1} 
\bar{\psi}\psi=\frac{c}{a(t)^3}.\ee
Self-interaction potential can be written as
$V=\sum_n\alpha_n\treug^{2n}$, where $\alpha_n$ and $n$ are constants. The 
potential $ V $ is considered as a scalar invariant.

\section{Reconstruction of solutions}

We consider our model for the case of power-law dependence of  scale factor on 
the time ($a(t)=a_0 t^n$) when the function $F(R)$ has a form $F(R) = r_0 R^p$.
Spinor field we select as (\ref{eqpsipsib1}) and  limit the potential to the 
first three terms
\bb  V(\bar\psi\psi)=\alpha_1
(\bar\psi\psi)^2+\alpha_2 (\bar\psi\psi)^4+\alpha_3
(\bar\psi\psi)^6. \label{usl2} \ee

\subsection{Model 1}

Let us consider the action in the following form ($F(R)=1$):
\bb S=\int{d^4\sqrt{-g}\{\frac{R}{2}+{\cal
L}_{\phi}+{\cal L}_D\}} . \ee

One can choose the Lagrangian of the scalar field as  (\ref{lagphi}),
we get the following expressions
\bea V(\phi) &=& -\frac{
   2 c (a^{15} m_f + a^{12} c \alpha_1 + a^6 c^3 \alpha_2 +
       c^5 \alpha_3)}{2 a^{18} r_0}+\nonumber\\
&+&\frac{ - 6 a^{16} a'^2 +
    a^{18} r_0 \phi'^2}{2 a^{18} r_0}\eea
\bea &&0=-\frac{2 a'^2}{a^2} + r_0 \phi'^2 +\\
&&+\frac{
  a^{15} c m_f + 2 a^{12} c^2 \alpha_1 + 4 a^6 c^4 \alpha_2 +
   6 c^6 \alpha_3 + 2 a^{17} a''}{a^{18}},\nonumber\eea
\bea  &&0=-2 a^{16} a'^3 + 
  2 a^{18} (a r_0 \phi' \phi'' + a^{(3)})+\\
&&3 a' (-2 (a^{12} c^2 \alpha_1 + 6 a^6 c^4
\alpha_2 + 15 c^6 \alpha_3) + a^{18} r_0 \phi'^2) .\nonumber\eea

Then, using that
$a(t)=a_0 t^n,\; \phi(t)=f_0 t^k,$
we get the following solutions:

1) $n=\frac {2}{3},\; \alpha_2=\alpha_3=0,\; m_f=\frac{4 a_0^3}{3
c},$\\
$k=-1,\; \alpha_1=-\frac{a_0^6f_0^2r_0}{2c^2},\; V(\phi)=0.$

2) $n=\frac {2}{3},\; \alpha_1=\alpha_3=0,\; m_f=\frac{4 a_0^3}{3
c},$\\
$k=-3,\; \alpha_2=-\frac{9 a_0^{12}f_0^2r_0}{4 c^4},\; V(\phi)=-\frac{9 
\phi^{8/3}}{4 f_0^{2/3}}=-\frac{9f_0^2}{t^8}.$

3)$n=\frac {2}{3},\; \alpha_1=\alpha_2=0,\; m_f=\frac{4 a_0^3}{3
c},$\\
$k=-5,\; \alpha_1=-\frac{25 a_0^{18}f_0^2r_0}{6 c^6},\; V(\phi)=-\frac{25 
\phi^{12/5}}{3 f_0^{2/5}}=-\frac{25 f_0^2}{3 t^{12}}.$

4)$n=\frac {1}{3},\; \alpha_2=\alpha_3=0,\;
m_f=-\frac{a_0^3f_0^2r_0}{4 c},$\\
$k=\frac{1}{2},\; \alpha_1=\frac{a_0^6}{3 c^2},\; 
V(\phi)=\frac{3f_0^4}{8\phi^2}=\frac{f_0^2}{8 t}.$

5)$n=\frac {1}{6},\; \alpha_1=\alpha_3=0,\; m_f=-\frac{9
a_0^3f_0^2r_0}{16 c},$\\
$k=\frac{3}{4},\; \alpha_2=\frac{a_0^{12}}{12 c^4},\; 
V(\phi)=\frac{9f_0^{8/3}}{32\phi^{2/3}}=\frac{9 f_0^2}{32 t^{1/2}}.$

6)$n=\frac {1}{9},\; \alpha_1=\alpha_2=0,\; m_f=-\frac{25
a_0^3f_0^2r_0}{36 c},$\\
$k=\frac{5}{6},\; \alpha_3=\frac{a_0^{18}}{27 c^6},\; V(\phi)=\frac{25 
f_0^{12/5}}{72 \phi^{2/5}}=\frac{25 f_0^2}{72 t^{1/3}}.$

We have several solutions that meet the decreased expansion. This situation is 
obvious, because the presence of of the spinor field leads to slower the 
universe expansion.

\subsection{Model 2}

Let us consider our model in the absence of a spinor field
\bb S=\int{d^4\sqrt{-g}\{\frac{R}{2}+f(R){\cal L}_{\phi}\}}.\ee

If the Lagrangian of the spinor field has the form (\ref{lagphi}) and
$$a(t)=a_0 t^n,\; \phi(t)=f_0 t^k,\; f(R)=r_0 R^p,$$
then we obtain the following solution
\begin{eqnarray} V(\phi)
&=&\frac{f_0^{2/p} p^2 (1 - 3 n + p) \phi^{2 - 2/p}}{2 (-1 + p)}=\nonumber\\
& =&\frac{f_0^2 p^2 (1 - 3 n + p) t^{-2 + 2 p}}{2 (-1 +
p)}\end{eqnarray}
$$ k = p,$$$$ r_0 = \frac{6 n^{1 - p} (-1 + 2 n) (-6 + 12 n)^{-p} (-1 +
p)}{ f_0^2 p^2 (3 - 6 n - p + 3 n p)}.$$ p and n  may be arbitrary, except 
$n=0,\;
n=1/2,\; p=0,\; n=\frac{p-3}{3(p-2)}.$

For this model, the $n$ can be arbitrary. If one selects the potential of the 
scalar field  to be zero, we obtain the limit for $n$
$$n=\frac{1+p}{3}.$$
The same solution was obtained in the review \cite{lit7} as realistic cosmology 
satisfying observational bounds and predicted by modified gravity.

\subsection{Model 3}

We now choose the action in its original form (\ref{act}), where the Lagrangian 
of the scalar field is
(\ref{lagphi}) and $a(t)=a_0 t^n,\;  f(R)=r_0 R^p.$














1.1)
$ V(\phi) = (434 - 38 \sqrt{73}) f_0^2 t^{1/6 (-5 - \sqrt{73})}
-$
\bea&&-
    \phi' \frac{(11 + 7 \sqrt{73}) \phi'}{
  (-19 + \sqrt{73})^2 (1 + a_0^2 t^{4/3})} -\\
  &&-   \phi' \frac{  3 a_0^2 t^{4/3} ((217 - 19 \sqrt{73})  \phi' + (
            52 \sqrt{73}-412) t  \phi''))}{
  (-19 + \sqrt{73})^2 (1 + a_0^2 t^{4/3})}.\nonumber\eea
Substituting $\phi(t)=f_0 t^k$ we get,
$$V(\phi)= -\frac{1}{144} (11 + 7 \sqrt{73}) f_0^{\frac{5}{2} - 
\frac{\sqrt{73}}{2}} \phi^{\frac{1}{2} (-1 + \sqrt{73})}=$$
  \bb =
-\frac{1}{144} (11 + 7 \sqrt{73}) f_0^2 t^{\frac{1}{6} (-17 -
\sqrt{73})},\ee
$$\alpha_2 = 0,\; \alpha_3 = 0,\; n = 2/3,\; m_f =\frac{4 a_0^3}{3 c},$$ $$ k = p 
-
1=-1.12867,$$
$\alpha_1 = -\frac{
   2^{1/6 - \sqrt{73}/6} 3^{1/12 (\sqrt{73}-31)} (19 \sqrt{73}-217) a_0^6 
f_0^2 r_0}{(-19 +
      \sqrt{73}) c^2}=$ $$=-\frac{0.27989 a_0^6 f_0^2 r_0}{c^2},\;
p = \frac{1}{12} (7 - \sqrt{73})=-0.128667.$$

1.2)
$V(\phi) = \frac{1}{ 4 (-1 + 2 p + 6 a_0^2 (-1 + p) p t^{2/3})}\times$\\
$\times((-1)^{-p} 
t^{-1 +2 p} (-e^{i p \pi} (f_0 + 2 f_0 p)^2 -$\\
$-
     2 (-1)^p t^{1 - 2 p} \phi' ((-1 - 2 p - 6 a_0^2 p (-2 + 3 p) t^{2/3}) \phi' 
+ 12 a_0^2 p t^{5/3} \phi''))),$
$$  V(\phi)= \frac{1}{8} f_0^{\frac{4}{1 + 2 p}} (1 + 2 p)^2 \phi^{\frac{-2 + 
4 p}{1 + 2 p}}= \frac{1}{8} f_0^2 (1 + 2 p)^2 t^{-1 + 2 p},$$
$$\alpha_2 = 0,\; \alpha_3 = 0,\; n = \frac{1}{3},$$$$ m_f =-\frac{2^{p-2}3^{-p} 
a_0^3 e^{i p \pi}f_0^2r_0(2p+1)^2}{c},$$$$ k = p
+ \frac{1}{2},\;\alpha_1 = \frac{a_0^6}{3 c^2}.$$

1.3)
$V(\phi) = \frac{1}{r_0} - \frac{49}{4} f_0^2 t^{3/2} + (\frac{9}{2} + 9 a_0^2 
t^{1/3}) \phi'^2 -
   12 a_0^2 t^{4/3} \phi' \phi'',$
   \bb V(\phi) = \frac{1}{r_0} + \frac{49}{32} f_0^2 
\left(\frac{\phi}{f_0}\right)^{6/7} = \frac{1}{r_0}+\frac{49}{32}f_0^2 
t^{3/2},\ee
$$\alpha_1 = 0,\; \alpha_3 = 0,\; n = \frac{1}{6},\; m_f =\frac{49 a_0^3 f_0^2 
r_0}{24 c},$$$$ k = p
+ \frac{3}{4},\;\alpha_2 = -\frac{a_0^{12}}{12 c^4},\;p=1.$$

1.4)
$V(\phi) = \frac{1}{r_0} - \frac{847}{36} f_0^2 t^{5/3} +
   \frac{15}{2} (1 + 2 a_0^2 t^{2/9}) \phi'^2 -
   18 a_0^2 t^{11/9} \phi' \phi'',$
    \bb V(\phi)= \frac{1}{r_0} + \frac{121}{72} f_0^{\frac{12}{11}} 
\phi^{\frac{10}{11}}= \frac{1}{r_0}+\frac{121}{72}f_0^2 t^{5/3},\ee
$$\alpha_1 = 0,\; \alpha_2 = 0,\; n =\frac{1}{9},\; m_f =\frac{847 a_0^3 f_0^2 
r_0}{486 c},$$$$ k = p
+ \frac{15}{18},\;\alpha_3 = -\frac{a_0^{18}}{27 c^6},\;p=1.$$

In this case, we see that many potentials allows us to find the explicit 
solutions. However, in several cases the expansion maybe decelerated. We dont 
discuss the details of the found solutions because it is known that several 
versions of them satisfy the observational bounds.




\section{Conclusions}

Thus, in our model we have a different types of behavior of the universe 
expansion. The presence of the spinor field leads to a slowing of the universe 
expansion, when the scale factor is positive and less than one ($n=1/9$, $1/6$, 
$1/3$ and $2/3$).
For the case of a free scalar field
if $n=2/3$ then we get the scalar field decreasing over time. Otherwise, the 
scalar field increases with time ($\phi \sim $ $t^{1/2}$, $t^{3/4}$ and 
$t^{5/6}$).

If we consider the model in the absence of a spinor field the situation is 
changing. The presence of non- minimal interaction allows to obtain solutions 
for any value of the degree in the scale factor. The degree of a scalar field 
is arbitrary.
We have only one restriction - $k=p$ (where $\phi=f_0 t^k$ and $F(R)=r_0 R^p$).

All the solutions we obtained for the power-law scalar field  ($\phi \sim 
t^k$). Choosing a different type of fields, such as logarithmic function of 
time, lead us to an equation without explicit solution.

Consider as an example the case of the scalar potential field set to $ 1/2 m 
\phi^2$. In this case, the equation (\ref{scal}) gets the  form
$$2 m \;t \;\phi + (3 n - 2 p) \dot{\phi} + t\ddot{\phi}=0.$$
The solution of this equation is
$$\phi= t^{\frac{1}{2}-\frac{3 n}{2}+p} 
(\text{BesselJ}\left[\frac{1}{2}-\frac{3 n}{2}+p,\sqrt{2} \sqrt{m} 
t\right] c_1+$$$$\text{BesselY}\left[\frac{1}{2}-\frac{3 n}{2}+p,\sqrt{2} \sqrt{m} 
t\right] c_2),$$
where $c_1$ and $c_2$ are constant, $\text{BesselJ}$ is the Bessel function of 
the first kind and $\text{BesselY}$ is the Bessel function of the second kind.
We see that in this case it would be difficult to check the compatibility of 
the solutions with the Einstein  equations. For this reason, we restricted 
ourselves to the power dependence of the scalar field on time.

If the degree of the scale factor is  positive, we  get  the quintessence-type 
universe. However, we can consider the case of a negative power. One can do a 
replacement $t\to t-t_s$ ($t_s$ is a constant) and  we obtain  the phantom 
universe with the singularity of the future such as the Big Rip.

We see that the presence of a  spinor field of specific type  does not permit 
the universe to expand with acceleration. The introduction of a scalar field 
does not change the situation.
However, in the absence of a spinor  field the  non-minimal interaction leads 
us to arbitrary powers of  the  scale factor and the scalar (in power form), 
but fixing potential and function $f(R)$.

\section*{Acknowledgments}

This work has been supported by  project 2.1839.2011 of Min. of Education and
Science (Russia) and LRSS project  224.2012.2 (Russia).


\end{document}